\documentclass{elsart}
\usepackage{graphicx}
\usepackage{amsfonts}
\usepackage{indentfirst}
\usepackage[small]{caption2}
\usepackage{longtable}
\usepackage{epsfig}

\def\MSbar{$\overline{\mathrm{MS}}\ $}

\def\Li#1#2{{\mathrm{Li}}_{#1}\left(#2\right)}

\def\ba{\begin{eqnarray}}
\def\ea{\end{eqnarray}}
\def\DD{{\mathcal D}}
\def\dd{{\mathrm d}}

\def\fun#1#2{\lower3.6pt\vbox{\baselineskip0pt\lineskip.9pt
  \ialign{$\mathsurround=0pt#1\hfil##\hfil$\crcr#2\crcr\sim\crcr}}}
\def\order#1{{\mathcal O}\left(#1\right)}

\def\gsim{\mathrel{\raise.3ex\hbox{$>$\kern-.75em\lower1ex\hbox{$\sim$}}}}
\def\lsim{\mathrel{\raise.3ex\hbox{$<$\kern-.75em\lower1ex\hbox{$\sim$}}}}
\begin{document}
\begin{frontmatter}

\title{QED collinear radiation factors in the next-to-leading logarithmic approximation}

\author{A.B. Arbuzov},
\author{E.S. Scherbakova}

\address{Bogoliubov Laboratory of Theoretical Physics, \\
JINR,\ Dubna, \ 141980 \ \  Russia }

\begin{abstract}
The effect of the collinear photon radiation by charged particles is considered
in the second order of the perturbation theory. Double and single photon 
radiation is evaluated. The corresponding radiation factors are obtained.
The QED renormalization group approach is exploited in the next-to-leading
order. The results are suited to perform a systematic 
treatment of the second order next-to-leading logarithmic radiative corrections 
to various processes either analytically or numerically.
\end{abstract}


\end{frontmatter}

%%%%%%%%%%%%%%%%%%%%%%%%%%%%%%%%%%%%%%%%%%%%%%%%%%%%%%%%%%%%%%%%%
\section{Introduction}  \label{sect_1}

The modern high energy physics experiments with advanced techniques
and high statistical require adequately precise theoretical predictions. 
Among various effects which have to be taken into account, 
QED radiative corrections (RC) give important contributions to the predictions. 
At high energies they are usually computed 
with help of the QED perturbation theory. But direct computations of higher
order QED corrections to complicated processes can be rather cumbersome. 
For this reason certain methods were developed to evaluate first the numerically most
important contributions. In particular, besides the expansion
in the powers of the fine structure constant $\alpha$, one can use an expansion in powers of 
the so-called large logarithm, $L=\ln(M^2/m^2)$, where $M$ is a large energy scale, 
and $m$ is a charged particle mass, $m\ll M$.

In this paper we present the derivation of a particular contribution of QED RC 
of the order $\order{\alpha^2L^{2,1}}$. 
It is well known, that the angular distribution of a 
photon emitted by a high--energy particle is peaked in the forward direction.
Moreover, it is easy to show starting from the matrix element, 
that a process with emission of a collinear photon can be represented in a factorized 
form~(see {\it e.g.} Ref.~\cite{Arbuzov:1997pj}). 
As usually the factorization appear if it is possible to separate the long-distance 
sub-process of collinear photon emission and the main short-distance
sub-process. In other words, we assume that the experimental conditions of the particle 
registration allow to neglect the effects of small changes of transverse momenta 
arising from emission of the photon at the small angle with respect to its parent 
particle: $\vartheta_\gamma\ll 1$.
So the cross section (or the decay width) of the process with hard collinear photon emission 
can be represented as a convolution of the radiation factor 
$R$ and the distribution of the radiation-less processes $\dd\hat{\sigma}$ (in example 
of the $2\to2$ type):
\ba
& & \dd\sigma[a(p_1) +b(p_2) \to c(q_1) + d(q_2) + \gamma(k\sim(1-z)p_1)] 
\nonumber \\ 
& & \qquad \qquad  \qquad 
=  \dd\hat\sigma[a(zp_1)+b(p_2) \to c(q_1) + d(q_2)]\otimes R_{\mathrm{H}}^{\mathrm{ISR}}(z), 
\\ 
& & \dd\sigma[a(p_1)+b(p_2) \to c(q_1)+d(q_2) + \gamma(k\sim(1-z)q_1)] 
\nonumber \\ 
& & \qquad \qquad  \qquad 
=  \dd\hat\sigma[a(p_1)+b(p_2) \to c(q_1)+d(q_2)] R_{\mathrm{H}}^{\mathrm{FSR}}(z), \nonumber
\ea
where $z=E'/E$ is the energy fraction of the particle emitted the 
photon, $E$ and $E'$ are the charged particle energy {\em before} and {\em after}
radiation of the photon, respectively. In the case of the final state radiation (FSR),
we observed the energy of particle $c$ being equal to $zq_1^0$, and we have a direct 
product of the two factors. 
In the case of the initial state radiation (ISR), we usually compute the kernel cross section
in the center-of-mass reference frame of particles $a(zp_1)$ and $b(p_2)$ and then 
perform a relativistic boost to the laboratory reference frame. 

This paper is organized as follows. In the next Section we re-call the known
results for the first order collinear radiation factors. In Sect.~\ref{sect_3} and in
Sect.~\ref{sect_4} we present our results for the second order radiation factors for
double and single hard photon radiation, respectively. Possible applications of the
results are discussed in Conclusions. In Appendix we give the explicit formulae for
QED splitting functions used in the derivation of the factors.

%%%%%%%%%%%%%%%%%%%%%%%%%%%%%%%%%%%%%%%%%%%%%%%%%%%%%%%%%%%%%%%%%
\section{The First Order Approximation} \label{sect_2}

The derivation of the collinear radiation factors due to an emission of a single hard photon 
in $\order{\alpha}$ can be found in Ref.~\cite{Arbuzov:1997pj}. The factors read
\ba \label{RH}
R_{\mathrm{H}}^{\mathrm{ISR}}(z) &=&\frac{\alpha}{2\pi} \biggl[\frac{1+z^2}{1-z}
          \biggl( \ln\frac{E^2}{m^2} - 1 + l_0\biggr) +1-z
+ \order{\frac{m^2}{E^2}} + \order{\vartheta_0^2}
\biggr], 
\\
R_{\mathrm{H}}^{\mathrm{FSR}}(z) &=&\frac{\alpha}{2\pi}\biggl[ \frac{1+z^2}{1-z}
          \biggl( \ln\frac{E^2}{m^2} - 1 + l_0 + 2 \ln z \biggr) +1-z
+ \order{\frac{m^2}{E^2}} + \order{\vartheta_0^2}
\biggr].
\ea
The mass of the particle $m$ is assumed to be small compared with the energy, and
terms suppressed by the factor $m^2/E^2$ are omitted. The photon emission angle with respect
to its parent particle is restricted by the condition
\ba
\vartheta_\gamma < \vartheta_0, \qquad \frac{m}{E} \ll \vartheta_0 \ll 1,
\qquad 
l_0 = \ln\frac{\vartheta_0^2}{4}\, .
\ea
The energy of the emitted photon is assumed to be above a certain threshold, 
$E_\gamma > \Delta E$. The parameters $\vartheta_0$ and $\Delta$ either might 
be related to concrete experimental conditions, or serve as auxiliary quantities. 
In the latter case they should cancel out after summing up the contributions due to 
emission of the collinear hard photons with the ones of non-collinear hard photons 
and of soft photons. These $\order{\alpha}$ radiation factors are universal and 
describe collinear single photon emission for various high--energy 
processes~\cite{Arbuzov:1997pj}.

%%%%%%%%%%%%%%%%%%%%%%%%%%%%%%%%%%%%%%%%%%%%%%%%%%%%%%%%%%%%%%%%%%%%%%%%%%%%%%%%%
\section{Double Hard Photon Radiation} \label{sect_3}

In paper~\cite{Arbuzov:1996qb} the effect of the double hard photon radiation
in the Bhabha scattering was considered. In particular the effect of the two
photon emission inside a collinear cone along the direction of motion of 
any of the 4 charged particles in this process was presented in a form being differential in 
the energy fraction of both the photons. So to get the collinear radiation factor,
we have just to integrate over one of the energy fractions keeping
their sum fixed. The lower limit of the integral over the photon energy fraction 
is chosen to be equal to the parameter $\Delta$ because both the photons should be hard
and have therefore energy above $\Delta E$.
In this way for the case of the initial state radiation we got
\ba\label{RHH}
R_{\mathrm{HH}}^{\mathrm{ISR}}(z) &=& \biggl(\frac{\alpha}{2\pi}\biggr)^2L\biggl\{ 
        (L+2l_0) \biggl( \frac{1+z^2}{1-z}(2\ln(1-z)- 2\ln\Delta-\ln z) 
+ \frac{1+z}{2}\ln z 
\nonumber \\
&-& 1 + z  \biggr) 
+  \frac{1+z^2}{1-z} \biggl(\ln^2 z + 2\ln z - 4\ln(1-z) + 4\ln\Delta \biggr)
\nonumber \\
&+& (1-z)\biggl(2 \ln(1-z) - 2\ln\Delta - \ln z + 3\biggr) + \frac{1+z}{2}\ln^2z \biggr\},
%L Rpmpmtxt = alph^2*ALME*( 
%         (ALME+2*ALTET)*( (1+z^2)/[1-z]*(2*DL1Z-2*ALD-DLZ) + (1+z)/2*DLZ - 1 + z )
%       + (1+z^2)/[1-z]*(2*DLZ - 4*DL1Z + 4*ALD + DLZ^2)
%       + (1-z)*( -2*ALD + 3 - DLZ + 2*DL1Z) + (1+z)/2*DLZ^2 
%       );
\ea
where $z$ is, as in Eq.~(\ref{RH}), the energy fraction of the charged particle
{\em after} the emission of the two photons.

The corresponding radiation factor for the final state radiation case can be obtained
from the ISR one by means of the Gribov--Lipatov relation:
\ba
R_{\mathrm{HH}}^{\mathrm{FSR}}(z) &=& 
            \left. - zR_{\mathrm{HH}}^{\mathrm{ISR}}\biggl(\frac{1}{z}\biggr)
               \right|_{\ln\Delta\to\ln\Delta-\ln z;\ l_0\to l_0 +2\ln z}
       =\biggl(\frac{\alpha}{2\pi}\biggr)^2L\biggl\{
        (L+2l_0) \biggl[ \frac{1+z^2}{1-z}
\nonumber \\
&\times& \biggl(2\ln(1-z)
- 2\ln\Delta+\ln z\biggr) 
+ \frac{1+z}{2}\ln z 
- 1 + z  \biggr] 
\nonumber \\
&+&  \frac{1+z^2}{1-z} \biggl( 5\ln^2 z - 2\ln z - 4\ln(1-z) + 4\ln\Delta + 8\ln z(\ln(1-z)-\ln\Delta)
 \biggr)
\nonumber \\
&+& (1-z)\biggl(2\ln(1-z) - 2\ln\Delta - 3\ln z + 3\biggr) + \frac{3(1+z)}{2}\ln^2z \biggr\},
%L Rqmqmtxt =  alph^2*ALME*( 
%         (ALME+2*ALTET)*( 2*(1+z^2)/[1-z]*(DL1Z-ALD+DLZ/2) + (1+z)/2*DLZ - 1 + z )
%       + (1+z^2)/[1-z]*( 5*DLZ^2 + 4*ALD*(1-2*DLZ) + 8*DL1Z*DLZ - 4*DL1Z - 2*DLZ )
%       + (1-z)*(3-2*ALD-3*DLZ+2*DL1Z) + 3/2*(1+z)*DLZ^2
%       );
\ea
Note that the additional interchanges in the above relation applied for 
$\ln\Delta$ and $l_0$ appear in our case from the crossing relations of the two 
channels with the given cuts on the energies of the soft photon and on the 
photon emission angle.

%%%%%%%%%%%%%%%%%%%%%%%%%%%%%%%%%%%%%%%%%%%%%%%%%%%%%%%%%%%%%%%%%%%%%%%%%%%%%%%%%
\section{Single Hard Photon Radiation} \label{sect_4}

We have to consider also the process of single hard photon emission accompanied by 
the one-loop virtual correction or by the emission of a soft photon. As concerns soft photon
radiation, its contribution does factorize with respect to the collinear hard photon emission:
\ba
\dd\sigma_{\mathrm{HS}} = R_{\mathrm{H}}\otimes \delta_{\mathrm{S}} \dd\sigma^{(0)}, 
\qquad \delta_{\mathrm{S}} = \frac{\dd\sigma^{(1)}_{\mathrm{Soft}}}{\dd\sigma^{(0)}}\, ,
\ea
where $\delta_{\mathrm{S}}$ is the one-loop soft photon radiation factor for the given
process, computed in the standard way~\cite{'tHooft:1978xw}.
This quantity has an infra-red divergence, which cancels out after
summation with the virtual loop contribution. 
And $\dd\sigma^{(0)}$ is the Born level cross section.

So we would like to get the radiation factor $R_{\mathrm{H(S+V)}}^{\mathrm{ISR}}(z)$,
where both the one-loop virtual correction and soft photon radiation are taken
into account. To find this radiation factor, we will exploit the known result
for the complete second order NLO QED corrections provided by the renormalization group
approach~\cite{Kuraev:1985hb,Berends:1987ab,Skrzypek:1992vk,Arbuzov:1999cq,Arbuzov:2002cn}. 
In analogy to QCD we can write the master formula for the corrected cross 
section {\it e.g.} for Bhabha scattering in the form (see Ref.~\cite{Arbuzov:2006mu}):
\ba
\label{master}
\dd \sigma &=& \int^{1}_{\bar{z}_1} \dd z_1 \int^{1}_{\bar{z}_2} \dd z_2 
          \DD^{\mathrm{str}}_{ee} (z_1) \DD^{\mathrm{str}}_{ee} (z_2)
		\left( \dd \sigma^{(0)} (z_1,z_2) + \dd \bar{\sigma}^{(1)} (z_1,z_2) 
+ \order{\alpha^2L^0}  \right) 
\nonumber \\
&\times& \int^{1}_{\bar{y}_1} \frac{\dd y_1}{Y_1} \int^{1}_{\bar{y}_2} \frac{\dd y_2}{Y_2}
\DD^{\mathrm{frg}}_{ee} (\frac{y_1}{Y_1}) \DD^{\mathrm{frg}}_{ee} (\frac{y_2}{Y_2}),
\ea
where $\dd\bar\sigma^{(1)}$ is the $\order{\alpha}$ correction to the massless 
%Bhabha 
scattering, calculated using the \MSbar scheme to subtract the %lepton 
mass singularities.  
The energy fractions of the incoming partons are $z_{1,2}$, and 
$Y_{1,2}$ are the energy fractions of the outgoing electron and positron.
$\DD_{ee}^{\mathrm{str(frg)}}$ are the structure (fragmentation) functions of an
electron. Here we consider only the photonic contributions to the non-singlet
part of these functions. The radiation factors corresponding to the collinear 
emission of light pairs were evaluated in Ref.~\cite{Arbuzov:1995vi}.
With help of the master formula we can find the most
important contributions reinforced by the large logarithm $L$ in radiative corrections
to a wide class of other processes as well.

We are going to drop the pair contributions, so we need here the pure photonic 
part of the non-singlet structure (fragmentation) functions for the initial (final)
state corrections. These functions describe the
probability to find a massless (massive) electron with energy fraction $z$ 
in the given massive (massless) electron.   
In our case with the next-to-leading accuracy we have
\ba  \label{Dee}
\DD_{ee}^{\mathrm{str,frg}} (z) &=& \delta(1-z)
+ \frac{\alpha}{2\pi}d^{(1)}(z,\mu_0,m_e)
+ \frac{\alpha}{2\pi}LP^{(0)}(z)
 \nonumber \\
&+& \biggl(\frac{\alpha}{2\pi}\biggr)^2
\biggl(\frac{1}{2}L^2P^{(0)}\otimes P^{(0)}(z)
+ LP^{(0)}\otimes d^{(1)}(z,\mu_0,m_e)
\\ \nonumber 
&&\qquad \qquad + LP^{(1,\gamma){\mathrm{str,frg}}}_{ee}(z) \biggr)
+ \order{\alpha^2 L^{0}, \alpha^3}.
\ea
%where the symbol "$\otimes$" means the standard convolution operation 
%(see {\it e.g.} Ref.~\cite{Arbuzov:2003ed}).
The difference between the functions appear
only due to the difference in the next-to-order splitting functions $P^{(1,\gamma)}$,
given in the Appendix together with the other relevant functions.
The modified minimal subtraction scheme \MSbar is used.
We have chosen the factorization scale equal to $E$, and the renormalization scale
$\mu_0$ will be taken equal to $m_e$. 
More details on the application of the approach to calculation
of second order next-to-leading QED corrections can be found in 
Refs.~\cite{Berends:1987ab,Arbuzov:2002cn}. 

Let us consider the $\order{\alpha^2L^{n}}$ $(n>0)$ radiative corrections to 
a given process, which are
related to at least one hard photon emission. They can be separated into four parts
according to their kinematics:
\ba \label{d2collinear}
\delta^{(2)\mathrm{NLO}}_{\mathrm{Hard}} = 
\delta^{(2)}_{\mathrm{HH(coll)}}+
\delta^{(2)}_{\mathrm{HH(s-coll)}}+
\delta^{(2)}_{\mathrm{(S+V)H(n-coll)}}+
\delta^{(2)}_{\mathrm{(S+V)H(coll)}},
\ea
where $\delta^{(2)}_{\mathrm{HH(coll)}}$ gives the contribution of double hard photon
emission considered in the previous section. The case when one of the photons is emitted
at large angle $(\vartheta_\gamma >\vartheta_0)$ and the other one is collinear
is denoted $\delta^{(2)}_{\mathrm{HH(s-coll)}}$, where ``s-coll'' means a semi-collinear 
kinematics, see Ref.~\cite{Arbuzov:1996qb} for details.
The term $\delta^{(2)}_{\mathrm{(S+V)H(n-coll)}}$ corresponds to single hard non-collinear 
$(\vartheta_\gamma >\vartheta_0)$ photon emission accompanied by the $\order{\alpha}$ soft and virtual 
photonic corrections. Note that since the non-collinear photon emission doesn't give 
rise to the large 
logarithm, we can keep in $\delta^{(2)}_{\mathrm{(S+V)H(n-coll)}}$ only the leading 
logarithmic terms in the sum of soft and virtual corrections. And the last term is the 
contribution that we are looking for: the one due to single hard collinear photon emission
accompanied by $\order{\alpha}$ soft and virtual corrections.

From the other hand, the same quantity can be found in the master formula~(\ref{master}):
\ba \label{d2master}
\delta^{(2)\mathrm{NLO}}_{\mathrm{Hard}} &=& 
\frac{\alpha}{2\pi} L P_{\Theta}^{(0)}\otimes  \dd\bar\sigma_\Theta^{(1)}+
\frac{\alpha}{2\pi} L P_{\Delta}^{(0)}\otimes  \dd\bar\sigma_\Theta^{(1)}+
\frac{\alpha}{2\pi} (L P_{\Theta}^{(0)} +d^{(1)}_{\Theta})\otimes \dd\bar\sigma_\Delta^{(1)} 
\nonumber \\
&+& \left(\frac{\alpha}{2\pi} \right)^2 \biggl(\frac{L^2}{2} P^{(0)}\otimes P^{(0)}
      +  L P^{(0)}\otimes d^{(1)} + L P^{(1,\gamma){\mathrm{str}}}\biggr)_{\Theta}\otimes\dd\sigma^{(0)},
\ea
where we leaved out the splitting functions arguments for short. 
Here $\dd\bar\sigma_\Theta^{(1)}$ is the contribution 
of single hard photon emission and $\dd\bar\sigma_\Delta^{(1)}$ is 
the soft-virtual contribution (in the $\overline{MS}$ scheme with massless electrons).
Lower indexes $\scriptstyle{\Theta}$ and $\scriptstyle\Delta$ 
mean here the parts of the corresponding functions related to hard and soft plus virtual 
radiation, respectively.
%, see the Appendix for some explicit definition.
Again we kept in the above equation only the terms reinforced by the large logarithm $L$.

Comparing the two expression~(\ref{d2collinear}) and (\ref{d2master}) we get 
\ba\label{dSVHcal}
&& \delta^{(2)}_{\mathrm{(S+V)H(coll)}} 
 = R_{\mathrm{H(S+V)}}^{\mathrm{ISR}}(z)\otimes \dd\hat{\sigma}(z) 
= \frac{\alpha}{2\pi} L P^{(0)}\otimes
\dd\bar\sigma^{(1)}_\Theta
+ \frac{\alpha}{2\pi} L P^{(0)}_\Theta\otimes\dd\bar\sigma_\Delta^{(1)}
\nonumber \\ && \qquad
+\left(\frac{\alpha}{2\pi} \right)^2 \biggl(\frac{L^2}{2} P^{(0)}\otimes P^{(0)}
       + L P^{(0)}\otimes d^{(1)} + L P^{(1,\gamma){\mathrm{str}}}\biggr)_{\Theta}\otimes\dd\sigma^{(0)}
\nonumber \\ && \qquad
- \left.\frac{\alpha}{2\pi}R^{(0)}_{\Theta}\otimes
\dd\sigma_\Theta^{(1)}\right|_{\vartheta_\gamma\ge\vartheta_0}
-\left.\frac{\alpha}{2\pi} L P^{(0)}_\Delta\otimes
\dd\sigma_\Theta^{(1)}\right|_{\vartheta_\gamma\ge\vartheta_0}
- \left(\frac{\alpha}{2\pi} \right)^2 
   R_{\mathrm{HH}}^{\mathrm{ISR}}\otimes\dd\sigma^{(0)}. 
\ea

The \MSbar subtraction leads to the following relations:
\ba\label{pod1}
\dd\bar\sigma_\Delta^{(1)} &=& \dd\sigma_{\mathrm{Soft}}^{(1)}+\dd\sigma_{\mathrm{Virt}}^{(1)}
-\frac{\alpha}{2\pi}(LP_{\Delta}^{(0)}+d^{(1)}_{\Delta})\dd\sigma^{(0)},\\ \label{pod2}
\dd\bar\sigma_\Theta^{(1)} &=& \dd\sigma^{(1)}_\Theta
-\frac{\alpha}{2\pi}(LP_{\Theta}^{(0)}+d^{(1)}_{\Theta})\otimes\dd\sigma^{(0)}.
\ea
Summing up the parts in (\ref{dSVHcal}) proportional to $\sigma_\Theta^{(1)}$ with help of 
(\ref{pod2}) we arrive at 
\ba\label{pod3}
\left.\frac{\alpha}{2\pi}L P^{(0)}_\Theta\otimes\dd\sigma_\Theta^{(1)}\right|_{\vartheta_\gamma<\vartheta_0}
=\frac{\alpha}{2\pi}L P^{(0)}_\Theta\otimes\frac{\alpha}{2\pi}R^{\mathrm{ISR}}_{\mathrm{H}}
\otimes\dd\sigma^{(0)}.
\ea

After substitution (\ref{pod1}) and (\ref{pod3}) to (\ref{dSVHcal}) we get the result
\ba
&& R_{\mathrm{H(S+V)}}^{\mathrm{ISR}}(z)\otimes \dd\hat{\sigma}(z)
= \left(\frac{\alpha}{2\pi}\right)^2 L P^{(0)}\otimes
    	R^{\mathrm{ISR}}_{\mathrm{H}}\otimes \dd\sigma^{(0)}
\nonumber \\ && \qquad
- \left(\frac{\alpha}{2\pi}\right)^2 L P^{(0)}\otimes 
      (L P^{(0)}_{\Theta}+d^{(1)}_\Theta)\otimes \dd\sigma^{(0)}
- \left(\frac{\alpha}{2\pi} \right)^2 R_{\mathrm{HH}}^{\mathrm{ISR}}\otimes\dd\sigma^{(0)}
\nonumber \\  && \qquad
+ \frac{\alpha}{2\pi} L P^{(0)}_\Theta\otimes
      \biggl(\dd\sigma_{\mathrm{Soft}}^{(1)}+\dd\sigma_{\mathrm{Virt}}^{(1)}
      -\frac{\alpha}{2\pi}(LP_{\Delta}^{(0)}+d^{(1)}_{\Delta})\dd\sigma^{(0)}\biggr)
\nonumber \\  && \qquad
+\left(\frac{\alpha}{2\pi} \right)^2 \biggl(\frac{L^2}{2} P^{(0)}\otimes P^{(0)}
	        + L P^{(0)}\otimes d^{(1)} 
+ L P^{(1,\gamma){\mathrm{str}}}\biggr)_{\Theta}\otimes\dd\sigma^{(0)}.
\ea
					 
Using the tables of convolution integrals~\cite{Arbuzov:2003ed} we obtain the 
answer for the ISR factor
\ba
&&  R_{\mathrm{H(S+V)}}^{\mathrm{ISR}}(z)\otimes\dd\hat{\sigma}(z)
=\delta_{\mathrm{(S+V)}}^{(1)}R_{\mathrm{H}}^{\mathrm{ISR}}(z)\otimes\dd\sigma^{(0)}(z)
\nonumber \\ && \qquad
+\biggl(\frac{\alpha}{2\pi}\biggr)^2L \biggl[ 2\frac{1+z^2}{1-z} \biggl( \Li{2}{1-z} 
- \ln(1-z) \ln z \biggr)
\nonumber \\ && \qquad
- (1+z)\ln^2z + (1-z) \ln z  + z \biggr]\otimes\dd\sigma^{(0)}(z), 
\\ \nonumber &&
\delta_{\mathrm{(S+V)}}^{(1)} = \frac{\dd\sigma_{\mathrm{Soft}}^{(1)}+\dd\sigma_{\mathrm{Virt}}^{(1)}}
{\dd\sigma^{(0)}}\, .
\ea

To get the final state radiation factor we use again the Gribov--Lipatov relation and get
\ba
&& R_{\mathrm{H(S+V)}}^{\mathrm{FSR}}(z)\dd\hat{\sigma}
=\delta_{\mathrm{(S+V)}}^{(1)}R_{\mathrm{H}}^{\mathrm{FSR}}(z)\dd\sigma^{(0)}
\nonumber \\ && \qquad
+ \biggl(\frac{\alpha}{2\pi}\biggr)^2L\biggl[ \frac{1+z^2}{1-z}  (  - 2\Li{2}{1-z} 
- 3\ln^2 z + 2\ln(1-z)\ln z )
\nonumber \\ && \qquad
 + (1+z) \ln^2 z - (1-z) \ln z  - 1
\biggr]\dd\sigma^{(0)}. 
\ea

%%%%%%%%%%%%%%%%%%%%%%%%%%%%%%%%%%%%%%%%%%%%%%%%%%%%%%%%%%%%%%%%%
\section{Conclusions} \label{sect_5}

In this way, we received the explicit expressions for the radiation 
factors, which describe hard collinear photon emission in the second 
order of the perturbation theory within the next-to-leading logarithmic 
approximation. These factors are universal. They can be used in analytic 
and numeric calculations of QED radiative corrections to a wide range of processes. 
In particular, we are going to implement 
them into the Monte Carlo event generators {\tt LABSMC}~\cite{Arbuzov:1999db}, 
{\tt SAMBHA}~\cite{Arbuzov:2004wp}, and {\tt MCGPJ}~\cite{Arbuzov:2005pt}
for several high energy processes. Our results can be 
exploited also to provide advanced theoretical predictions for experimental
observables with so-called tagged photons, when hard photons emitted at 
zero (small) angles with respect to colliding charged particles are 
detected~\cite{Krasny:1991hd,Arbuzov:1998te,Anlauf:1998fg}.

%%%%%%%%%%%%%%%%%%%%%%%%%%%%%%%%%%%%%%%%%%%%%%%%%%%%%%%%%%%%%%%%%

\ack{
We are grateful to
E.~Kuraev  
for discussions.
This work was supported by 
the RFBR grant 07-02-00932 
and 
the INTAS grant 05-1000008-8328.
One of us (A.A.) thanks also the grant of the President RF
(Scientific Schools 5332.2006).
}

%%%%%%%%%%%%%%%%%%%%%%%%%%%%%%%%%%%%%%%%%%%%%%%%%%%%%%%%%%%%%%%%%
\section*{Appendix. Explicit Formulae for QED Splitting Functions}
%\vskip 20.0pt
\setcounter{equation}{0}
\renewcommand{\theequation}{A.\arabic{equation}}

The QED splitting functions corresponding to photonic corrections
in the leading logarithmic approximation in the first and second orders read
\ba
&& P^{(0)}_{ee}(z) = \left[\frac{1+z^2}{1-z}\right]_+ 
= \lim_{\Delta\to0}\biggl\{ \delta(1-z) P^{(0)}_{\Delta}
+\Theta(1-z-\Delta) P^{(0)}_{\Theta}(z)\biggr\},
\nonumber \\
&& P^{(0)}_{\Delta} = 2\ln\Delta+\frac{3}{2}, \qquad
P^{(0)}_{\Theta}(z) = \frac{1+z^2}{1-z}.
\\ \nonumber 
&& P^{(0)}_{ee}\otimes P^{(0)}_{ee}(z) =
\lim_{\Delta\to0}\biggl\{ \delta(1-z)\biggl[\biggl(2\ln\Delta + \frac{3}{2}\biggr)^2
 - 4\zeta(2)\biggr] 
\nonumber 
\\ \nonumber && \qquad
+ \Theta(1-z-\Delta)
2\biggl[ \frac{1+z^2}{1-z}\biggl(
2\ln(1-z) - \ln z + \frac{3}{2} \biggr) + \frac{1+z}{2}\ln z - 1 + z \biggr],
\ea
where symbols $\delta$ and $\Theta$ denote the Dirac $\delta$-function
and the step function, respectively.

The space-like (ISR) and time-like (FSR) next-to-leading terms of the QED 
splitting functions for photonic
corrections can be cast as \cite{Berends:1987ab,Arbuzov:2002cn}
\ba
&& P^{(1,\gamma){\mathrm{str}}}_{ee}(z) = 
\delta(1-z)\biggl( \frac{3}{8} - 3\zeta_2 + 6\zeta_3 \biggr)
+ \frac{1+z^2}{1-z}\biggl( - 2\ln z \ln(1-z) 
\nonumber \\ && \quad
+ \ln^2z + 2\Li{2}{1-z} \biggr)
- \frac{1}{2}(1+z)\ln^2z  + 2\ln z - 2z + 3,
\nonumber \\ 
&& P^{(1,\gamma){\mathrm{frg}}}_{ee}(z) =
\delta(1-z)\biggl( \frac{3}{8} - 3\zeta_2 + 6\zeta_3 \biggr)
+ \frac{1+z^2}{1-z}\biggl( 2\ln z \ln(1-z)
\nonumber \\ && \quad
- 2\ln^2z - 2\Li{2}{1-z} \biggr)
+ \frac{1}{2}(1+z)\ln^2z
+ 2z\ln z - 3z + 2.
\ea
In the next-to-leading calculations we need also the initial condition
for the structure and fragmentation functions at a certain scale $\mu_0$:
\ba
d^{(1)}(z,\mu_{0},m_e) &\equiv& d^{(1)}(z) = \biggl[ \frac{1+z^2}{1-z}
\biggl( \ln\frac{\mu_{0}^2}{m_e^2} - 2\ln(1-z) - 1 \biggr) 
\biggr]_+.
\ea

The dilogarithm and the Riemann zeta-function are defined as usually:
\ba
\Li{2}{x} = \int_0^1\dd y \frac{\ln(1-xy)}{y}\, , \qquad
\zeta(n) = \sum\limits_{k=1}^{\infty}\frac{1}{k^n}\, .
\ea

%-------------------------Bibliography-------------------


\begin{thebibliography}{99}

\bibitem{Arbuzov:1997pj}
A.B.~Arbuzov, G.V.~Fedotovich, E.A.~Kuraev,
N.P.~Merenkov, V.D.~Rushai, L.~Trentadue, 
%{\em Large Angle QED Processes at $e^+e^-$ colliders
%  at energies below 3 GeV,\/} \\
JHEP {\bf 10} (1997) 001.
%hep--ph/9702262, 21pp.
%%CITATION = HEP-PH 9702262;%%

\bibitem{Arbuzov:1996qb}
A.B.~Arbuzov, V.A.~Astakhov, E.A.~Kuraev, N.P.~Merenkov,
L.~Trentadue, E.V.~Zemlyanaya, 
%{\em Emission of Two Hard Photons in Large--Angle Bhabha Scattering,\/}
%Preprint JINR, Dubna, E2--95--460, 1995, 12pp.; hep--ph/9610228; \\
Nucl. Phys. {\bf B 483} (1997) 83.
%%CITATION = HEP-PH 9610228;%%

\bibitem{'tHooft:1978xw}
  G.~'t Hooft and M.J.G.~Veltman,
  %``Scalar One Loop Integrals,''
  Nucl. Phys. B {\bf 153} (1979) 365.
  %%CITATION = NUPHA,B153,365;%%

\bibitem{Kuraev:1985hb}
  E.A.~Kuraev and V.S.~Fadin,
  %``On Radiative Corrections To E+ E- Single Photon Annihilation At
  %High-Energy,''
  Sov.\ J.\ Nucl.\ Phys.\  {\bf 41} (1985) 466.
  %[Yad.\ Fiz.\  {\bf 41} (1985) 733].
  %%CITATION = SJNCA,41,466;%%

\bibitem{Berends:1987ab}
  F.A.~Berends, W.L.~van Neerven and G.J.H.~Burgers,
  %``Higher Order Radiative Corrections At Lep Energies,''
  Nucl.\ Phys.\ B {\bf 297} (1988) 429
  [Erratum-ibid.\ B {\bf 304} (1988) 921].
  %%CITATION = NUPHA,B297,429;%%

\bibitem{Skrzypek:1992vk}
  M.~Skrzypek,
  %``Leading logarithmic calculations of QED corrections at LEP,''
  Acta Phys.\ Polon.\ B {\bf 23} (1992) 135.
  %%CITATION = APPOA,B23,135;%%

\bibitem{Arbuzov:1999cq}
A.B.~Arbuzov, 
%{\em Non-singlet splitting functions in QED, \/} \\
Phys. Lett. {\bf B 470} (1999) 252.
%[hep--ph/9908361].
%%CITATION = HEP-PH 9908361;%%

\bibitem{Arbuzov:2002cn}
A.~Arbuzov, K.~Melnikov, 
%{\em ${\mathcal O}(\alpha^2 \ln(m_\mu/m_e))$ Corrections to Electron Energy
%Spectrum in Muon Decay,\/} \\
Phys.\ Rev.\ D {\bf 66} (2002) 093003.
%[hep-ph/0205172].
%%CITATION = HEP-PH 0205172;%%

\bibitem{Arbuzov:2006mu}
  A.B.~Arbuzov and E.S.~Scherbakova,
  %``Next-to-leading order corrections to Bhabha scattering in  renormalization
  %group approach. I: Soft and virtual photonic  contributions,''
  JETP Lett.\  {\bf 83} (2006) 427.
  %[arXiv:hep-ph/0602119].
  %%CITATION = JTPLA,83,427;%%

\bibitem{Arbuzov:1995vi}
  A.B.~Arbuzov, E.A.~Kuraev, N.P.~Merenkov and L.~Trentadue,
  %``Hard pair production in large angle Bhabha scattering,''
  Nucl.\ Phys.\  B {\bf 474} (1996) 271.
  %%CITATION = NUPHA,B474,271;%%

\bibitem{Arbuzov:2003ed}
  A.B.~Arbuzov,
  {\em Tables of convolution integrals,\/}
  hep-ph/0304063.
  %%CITATION = HEP-PH 0304063;%%

\bibitem{Arbuzov:1999db}
  A.B.~Arbuzov,
  {\em LABSMC: Monte Carlo event generator for large-angle Bhabha scattering,\/}
  hep-ph/9907298.
  %%CITATION = HEP-PH 9907298;%%

\bibitem{Arbuzov:2004wp}
A.B.~Arbuzov, D.~Haidt, C.~Matteuzzi, M.~Paganoni and L.~Trentadue, 
%{\em The running of the electromagnetic coupling alpha in small-angle Bhabha
%scattering,\/} \\
Eur.\ Phys.\ J.\ C {\bf 34} (2004) 267.
%[hep-ph/0402211].
%%CITATION = HEP-PH 0402211;%%

\bibitem{Arbuzov:2005pt}
A.B.~Arbuzov, G.V.~Fedotovich, F.V.~Ignatov, E.A.~Kuraev 
    and A.L.~Sibidanov, 
%{\em Monte-Carlo generator for e+e- annihilation into lepton and hadron 
%     pairs with precise radiative corrections, \/} \\
Eur. Phys. J. C {\bf 46} (2006) 689.
%[DOI: 10.1140/epjc/s2006-02532-8],[hep-ph/0504233].
%%CITATION = HEP-PH 0504233;%%

\bibitem{Krasny:1991hd}
  M.W.~Krasny, W.~Placzek and H.~Spiesberger,
  %``Determination of the longitudinal structure function at HERA from radiative
  %events,''
  Z.\ Phys.\  C {\bf 53} (1992) 687.
  %%CITATION = ZEPYA,C53,687;%%

\bibitem{Arbuzov:1998te}
  A.B.~Arbuzov, E.A.~Kuraev, N.P.~Merenkov and L.~Trentadue,
  %``Hadronic cross-sections in electron positron annihilation with tagged
  %photon,''
  JHEP {\bf 9812} (1998) 009.
  %[arXiv:hep-ph/9804430].
  %%CITATION = JHEPA,9812,009;%%

\bibitem{Anlauf:1998fg}
  H.~Anlauf, A.B.~Arbuzov, E.A.~Kuraev and N.P.~Merenkov,
  %``Tagged photons in DIS with next-to-leading accuracy,''
  JHEP {\bf 9810} (1998) 013.
  %[arXiv:hep-ph/9805384].
  %%CITATION = JHEPA,9810,013;%%

\end{thebibliography}
\end{document}